# Polarization Enhancement of terahertz radiation generated by intrinsic Josephson junctions in a truncated edge square $Bi_2Sr_2CaCu_2O_{8+\delta}$ mesa.

A. Elarabi[*][a], Y. Yoshioka[a], M. Tsujimoto[a], Y. Nakagawa[a], I. Kakeya[a]

[a]Department of Electronic Science and Engineering, Kyoto University

## 1. Introduction

In the last decade, great advances have been made in the field of Terahertz (THz) continuous–wave sources. THz radiation sources and methods have increased in number and improved in efficiency and power [1]. One of the most recently discovered THz sources is built from the high-Tc superconductor $Bi_2Sr_2CaCu_2O_{8+\delta}$(BSCCO) [2,3]. Utilizing the ac Josephson effect, the radiation in BSCCO based emitters is generated by an oscillating current in the intrinsic Josephson junctions (IJJs). The IJJs are formed as superconducting $CuO_2$ layers, stacked over insulating Bi-Sr-O layers along the c-axis of a mesa structure. Electromagnetic waves in general can be classified into three main classifications according to their polarization, linear, elliptical, and circular. However, only linearly polarized THz radiation was observed from BSCCO emitters so far [4]. Many new experiments and applications could be proposed if the polarization can be controlled in such emitters. Examples may include vibrational circular dichroism (VCD) spectroscopy, high speed communications and THz continuous–wave polarization imaging.

According to previous studies [5][6], the mesa structure resembles a microstrip patch antenna, therefore, Antenna Theory can be used to analyse the measured radiation properties, and — as in our study — predict it. The excitation of resonant modes plays a vital role in the emission of THz radiation from BSCCO mesa. Two spatially orthogonal resonant modes, which are equal in amplitude and differ in phase by 90 degrees, have to be excited in the mesa to observe a circularly polarized (CP) radiation. These two modes are resonant along the diagonal axis of the mesa, and can be controlled by modifying the areas of the truncated part and square space S of the mesa as described by these relations [7]:

$$f_{0r} = \frac{c_0}{2nw}, \qquad f_a = f_{0r}, \quad f_b = f_{0r}\left(1 + \frac{2\Delta s}{S}\right)$$

Where $c_0$, $n$, $w$, are the speed of light in vacuum, the refractive index, and the width of square mesa, respectively. The main design parameters for the truncated edge square mesa that has a side length $A$ are, $a_1$ the length of the truncated edge, and $a_2$ the length of the uncut part as shown in Figure 1, where $A=a_1+a_2$, $S = A^2$ and $\Delta s = a_1^2$. Note that $f_{0r}$, $f_a$, $f_b$ are the resonant frequency of the uncut patch, and the frequencies of the two new orthogonal modes of truncated patch, respectively. The condition for CP radiation occurs at the arithmetic mean value of these two resonant frequencies.

In this paper, the polarization characteristics of a truncated corner mesa design are introduced. This novel structure was used in another experiment, to generate an elliptically polarization radiation from a BSCCO THz emitter, which was the first observation of a non-linear polarization from a BSCCO THz source. The preliminary data in the experiment showed a highly elliptical polarization with an axial ratio approaching to 4 dB in the frequency range predicted by the cavity model. The details will be described elsewhere [8]. To enhance the polarization, an electromagnetic simulation (EM) is used to determine the axial ratio (AR) of emitted waves. Then by performing a parametric study, the enhanced axial ratios is observed and compared to the measured ones.

---

[*] Corresponding author. Tel.: +81-75-383-2271; fax: +81-75-383-2270.
E-mail address: asemelarabi@sk.kuee.kyoto-u.ac.jp

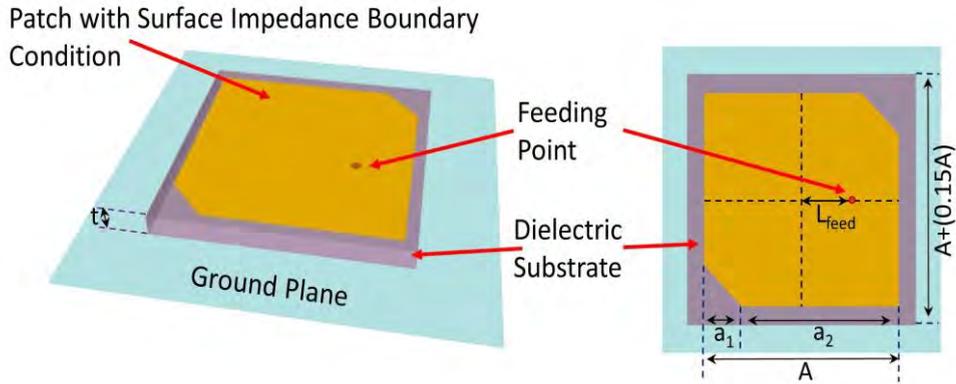

Fig. 1. A schematic view of the simulation model used in this study. The figure in the right shows a top view for the model.

## 2. Simulation Method

A commercial EM simulator named High Frequency Structural Simulator (HFSS) from ANSYS, Inc. was used to perform the simulation in this study. Using finite-element method (FEM), HFSS solves Maxwell's equation in 3D to find the required electromagnetic characteristics of a model. A similar method was used by Tsujimoto et al [9] using Sonnet EM simulator to identify the cavity modes of a nearly square mesa structure. Losses due to the frequency dependent skin effect was accounted for by applying a frequency dependant impedance boundary condition [10,11].The complex surface impedance $Z_s = R_s+iX_s$ , where $R_s$ is the surface resistance, and $X_s$ is the surface reactance, was found to have a significant effect on the electromagnetic properties of the mesa. The real part of the impedance Rs corresponds to the density of the unpaired charge carriers i.e., the conduction loss. In the other hand the imaginary part $X_s$ is associated with the density of the paired charge carriers, which relates to the penetration depth $\lambda L(T)$ ,and also describes the frequency dependent phase shift of the conductor. Layered superconductors like BSCCO contain very thin layers of superconductors separated by thicker dielectrics. Due to this layered nature, the current value and nature in the c-axis and ab-plane are quite different, where in the ab-plane it is similar to the bulk homogeneous superconductors, it can be described in the local London limits [12]. To estimate the surface impedance in this case, a self-contained computer code proposed by Zimmermann et al [11,13] was used. Despite being very important to explain the EM emission from IJJs, in this paper we didn't take IJJs synchronization into account. That being said, the simulation method used in here shows a good agreement with the experiments [8],and might help to clarify some radiation characteristics and improve the polarization.

The model used in the simulation is shown in Figure 1 ,in which two thin layers of perfect conductors are set over and beneath a dielectric box with a dielectric constant of $\epsilon = 17.6$ which is a typical value of an emitting BSCCO material [9]. The dimensions for the simulation model used in here, are correlated with those of a typically emitting square mesa structure [14].

## 3. Results and Discussions

Using the method mentioned earlier, the frequency dependent real part of the surface impedance was determined to be in the range $R_s$ = 0.159 - 0.15906 Ω in a frequency range of $f$ = 420 - 440 GHz. The imaginary part of the surface impedance was in the range $X_s$ = 1.34 -1.4 Ω within the same frequency range. The material parameters used to perform the simulation are of a typical BSCCO THz emitting mesa, in which the transition temperature $T_c$ = 82 K, the emission temperature $T_b$ = 40 K, the mean free path l = 10 nm, the London penetration depth $\lambda_L$ = 400 nm, and the coherence length $\xi_0$ = 0.1 nm. In general, the radiation characteristics acquired by applying the surface impedance boundary condition as performed in this study are in agreement with the experimental study, in contrary to the simulation done without applying it, which was not correlated to experimental results. The polarization direction in simulation was found to be a left hand polarization, where in the experiments it was not determined.

To enhance the polarization in similarly shaped mesa structures, two parameters were considered in the simulation, $a_1$ and $a_2$. At first $a_2$ was set to a fixed value of 68.7 μm , and the $a_1$'s value was incremented by 1 μm in every next run ,starting with $a_1$ = 5 μm. In figure 2 (b), the values for axial ratio (AR) for two values of $a_1$ swept in the frequency range of $f$ = 420 - 490 GHz as determined from the simulation is shown. The red line represents the AR at $a_1$ = 13 μm. At this value of $a_1$, the AR was elliptically polarized with a peak at $f$ = 440 GHz, though its value didn't exceed 5.3 dB .An enhanced value for the AR was found when $a_1$ was set to 9 μm, as shown by a black line in the figure. The minimum AR at this $a_1$ value was 0.77 dB at a frequency of 459 GHz, which is about 85.5% enhancement. At the minimum AR, the ratio between the truncated length $a_1$ to the uncut part length $a_2$ was at 13.1%. By decreasing the value of $a_1$ from 13 μm to 9 μm, the AR peak frequency has shifted by 19 GHz to the right (increased), yet it is still within the range attainable by experimentally produced IJJs emitters. The axial ratio band width (ARBW) at 3 dB was improved from 0 to 4 GHz (460.6 GHz - 456.6 GHz). In figure 2 (a), the scattering parameter is presented using the same colours explained for the two values of $a_1$. The impedance bandwidth (ZWB) at -10 dB was also improved from 0 to 2.8 GHz (465.5 GHz - 462.7 GHz).

Using a similar method, the second parameter $a_2$ was altered while the parameter $a_1$ was fixed to 13 μm. The results are shown in figure 2 (c)(d), where in figure 2 (d) the axial ratios for the values $a_2$ = 68 μm, and 94 μm are illustrated by a red line, and a black line, respectively. This time by increasing the $a_2$ value, an enhancement of 93.6% is realized, where the minimum AR was 5.6 dB at the first value of $a_2$, and 0.36 dB after enhancement. The frequency shift in AR peak was much larger than before, where it has shifted to the left (decreased) by 113 GHz from 444 GHz to 331 GHz. Comparing to the last parameter ratio value, a relatively similar $a_1$ to $a_2$ ratio of 13.8% was required to achieve the minimum AR. The ARBW has improved from 0 GHz in the case of $a_2$ = 13 μm in the simulation to 3.62 GHz (332.92GHz - 329.3GHz). Figure 2 (c) shows the scattering parameter values in the same frequency range. A very large frequency shift and peak difference values are also observed in it, though the peaks did not pass the -10 dB range.

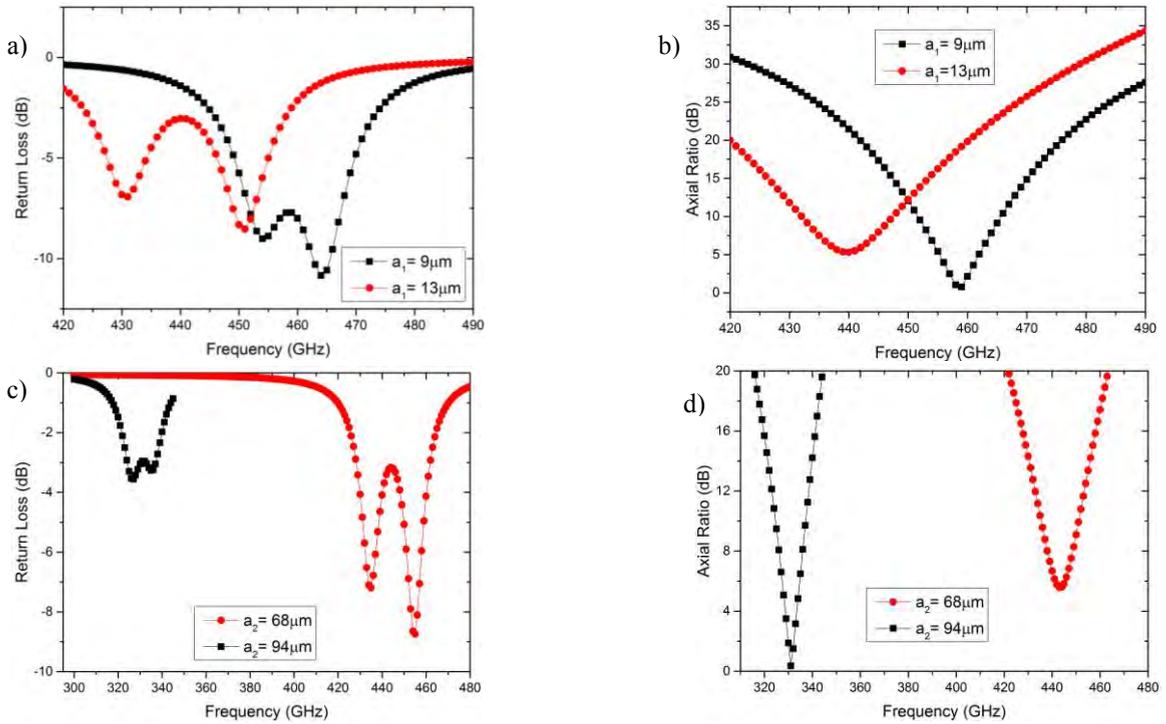

Fig. 2 (a) scattering parameter vs. frequency in the case of $a_2$=68.7 μm , $a_1$ = 9, 13 μm ; (b) the axial ratio for the mentioned values of $a_2$, $a_1$; (c) scattering parameter when $a_1$ = 13 μm, $a_2$ = 68, 94 μm (d) the axial ratio in this case.

## 4. Conclusion

In this study, the polarization characteristics of a truncated edge square mesa are numerically simulated using HFSS and impedance boundary condition. By using a parametric study approach, an enhancement for the polarization is realized, and discussed.

## Acknowledgements

The author would like to thank Mr. F.Tubbal from University of Wollongong in Australia for the technical support.